\documentclass[usegraphix,usenatbib]{mn2e}

\include{defn}

\usepackage{amsmath}
\usepackage{amssymb}
\usepackage{times}
\usepackage{graphicx}
\usepackage{subfigure}
\usepackage{epsfig}
\usepackage{txfonts}

\begin{document}

\pagenumbering{arabic}

\title[Hamiltonians of Spherical Galaxies in Action-Angle Coordinates]
  {Hamiltonians of Spherical Galaxies in Action-Angle Coordinates}

\author[Williams, Evans \& Bowden]
  {A.A. Williams$^1$\thanks{E-mail: aamw3,nwe,adb61@ast.cam.ac.uk},
   N.W. Evans$^1$,  A.D. Bowden$^1$
 \medskip
 \\$^1$Institute of Astronomy, University of Cambridge, Madingley Road,
       Cambridge, CB3 0HA, UK}

\maketitle

\begin{abstract}
We present a simple formula for the Hamiltonian in terms of the
actions for spherically symmetric, scale-free potentials. The
Hamiltonian is a power-law or logarithmic function of a linear
combination of the actions. Our expression reduces to the well-known
results for the familiar cases of the harmonic oscillator and the
Kepler potential. For other power-laws, as well as for the singular
isothermal sphere, it is exact for the radial and circular orbits, and
very accurate for general orbits. Numerical tests show that the
errors are always very small, with mean errors across a grid of
actions always $<1$ percent and maximum errors $<2.5$ percent. Simple
first-order corrections can reduce mean errors to $<0.6$ percent and
maximum errors to $<1$ percent.

We use our new result to show that: [1] the misalignment angle between
debris in a stream and a progenitor is always very nearly zero in
spherical scale-free potentials, demonstrating that streams can
sometimes be well-approximated by orbits, [2] the effects of an
adiabatic change in the stellar density profile in the inner region of
a galaxy weaken any existing $1/r$ dark matter density cusp, which is
reduced to $1/r^{1/3}$. More generally, we derive the full range of
adiabatic cusp transformations and show that a $1/r^{\gamma_0}$
density cusp may be changed to $1/r^{\gamma_1}$ only if $\gamma_0 /(4
- \gamma_0) \le \gamma_1 \le (9- 2\gamma_0)/(4-\gamma_0)$. {\it It follows
that adiabatic transformations can never completely erase a dark
matter cusp.}
\end{abstract}


\section{INTRODUCTION}
Action-angle coordinates are an extremely useful set of coordinates
with which to describe a stellar system. Actions are integrals of
motion given by
\begin{equation}
J_{i} = \dfrac{1}{2\pi}\oint_{\gamma_{i}} \boldsymbol{p}.\mathrm{d}\boldsymbol{q},
\end{equation}
where $\gamma_{i}$ is a closed path around an orbital torus, and
$(\boldsymbol{p},\boldsymbol{q})$ are canonically conjugate
phase-space coordinates. Since actions are integrals of motion, they
satisfy $\dot{J_{i}}=0$, greatly simplifying Hamilton's equations for
the canonically conjugate angle variables. These follow the trivial
time evolution
\begin{equation}
\theta_{i}(t)=\theta_{i}(0)+\Omega_{i}t,\qquad\qquad
\Omega_{i} = \frac{\partial H}{\partial J_{i}}.
\end{equation}
Already, we can see the appeal of the action-angle coordinate
space. In the physical phase-space $(\boldsymbol{x},\boldsymbol{v})$,
each of the six phase-space coordinates generally evolves in a
complicated fashion. In $(\boldsymbol{J},\boldsymbol{\theta})$ space,
three of the coordinates are constant and the other three evolve
linearly with time. Action-angle space is the natural arena in which
to view stellar orbits.

The benefit of action-angles extends beyond their simple time
evolution, however. Despite the fact that galaxies are never truly in
a steady state, equilibrium models are hugely important tools in
developing our understanding of galaxies. Using equilibrium models, we
can infer the gravitational potential of a galaxy and predict all
observable quantities.  Under most circumstances, we are justified in
assuming that a galaxy is always close to an equilibrium state, and we
can model non-equilibrium effects such as accretion or stellar
evolution as small perturbations to the system.

Given the assumption of dynamical equilibrium, the most efficient way
to proceed is via the use of the Jeans theorem (see \citealt{Bi08}),
which tells us that the phase-space distribution function (hereafter
DF) of a galaxy or stellar system in equilibrium may be written as a
function of just three isolating integrals. This reduces the
dimensionality of the space under consideration from six to three, and
so invoking the Jeans theorem leads to an elegant simplification of
the modelling problem. Since actions are adiabatic invariants, if we
choose to write the DF as a function of the actions, its functional
form is preserved under slow changes to the galaxy. This property can
be exploited to model various processes in galaxy evolution.  For
example, \citet{Go84} showed that the adiabatic growth of mass at the
centre of a stellar system causes the velocity dispersion to become
tangentially distended, a process invoked by \citet{Ag14} to explain
the kinematics of M87's globular cluster populations.  Similarly,
\citet{Ka14} recently modelled the adiabatic compression of dark
matter halos in galaxies from the THINGS survey to reproduce their
rotation curves. Another example is provided by \citet{Go99}, who
showed that the adiabatic growth of a black hole can cause a dark
matter spike at the center of a cusped dark halo.

Unfortunately, although action-angles are very useful when known, they
are not usually available analytically. In fact, \citet{Ev90} showed
that the most general potential in which the actions are available as
elementary functions is the isochrone (which reduces to the Kepler and
harmonic potentials in limiting cases). Recent efforts by
\citet{Bi12}, \citet{Bo14} and \citet{Sa14} mean that actions can now
be numerically recovered in many potentials, but analytical
approximations are still sparse. Though numerical methods are
ultimately the most widely applicable, analytical approximations are
important insofar as they give us valuable insight into the properties
of a system, not to mention drastically reduced computational time
when they are sufficiently accurate.

In this paper, we present a new method for finding accurate
approximations to the Hamiltonians of spherically symmetric,
scale-free systems and apply it to general power-law potentials and
the isothermal sphere.  We then use the method in two applications:
finding the misalignment angle of a stellar stream, and calculating
the effect of a slowly changing central stellar density profile on the
slope of a dark matter cusp.

\section{Method}
In a spherical potential, the radial action is given by
\begin{equation} \label{eq:gensol}
J_{r} = \dfrac{1}{2\pi}\oint\sqrt{2E-2\Phi(r)-L^{2}/r^{2}}\mathrm{d}r.
\end{equation}
The angular actions are related to the angular momentum components via
$J_\phi = |L_z|$ and $J_\theta = L - |L_z|$, where $L= J_\theta +
J_\phi$ is the total angular momentum and $L_{z}$ is the angular
momentum about the $z$-axis. The Hamiltonian can then be regarded as a
function of just two quantities: $H\equiv H(L,J_{r})$. The most
general method of finding $H(L,J_{r})$ is to solve the integral
equation (\ref{eq:gensol}) for $E$ but, as previously mentioned, this
cannot be done for the general case.

We now present a prescription for finding approximations
$\mathcal{H}(L,J_{r})$ to the exact Hamiltonian $H(L,J_{r})$ of a
spherical, scale-free potential $\Phi(r)$.  In terms of the usual
phase-space coordinates, the Hamiltonian is given by
\begin{equation} \label{eq:hxv}
H(r,\boldsymbol{v}) = \frac{1}{2}\boldsymbol{v}^{2} + \Phi(r).
\end{equation}
We now look to find an approximate transformation by considering only
the two most basic types of orbit: radial and circular orbits. For the
radial orbits, we must evaluate
\begin{equation}
J_{r} = \dfrac{1}{2\pi}\oint\sqrt{2E(0,J_{r})-2\Phi(r)}\mathrm{d}r
\label{eq:Jr}
\end{equation}
and invert the resulting expression for $E(0,J_{r})$. The energy of
the circular orbits is given by the system of simultaneous equations
\begin{equation}
L(r) = rv_{\rm c}(r), \qquad\qquad E(L,0) = \frac{1}{2}v_{\rm c}(r)^{2} +
\Phi(r),
\label{eq:L}
\end{equation}
where $v_{\rm c}(r)$ is the circular speed. For a scale-free
potential, simply as a consequence of dimensional analysis, both
results will have the same functional form
\begin{equation}
E(J_{i},J_{j}=0) = f(a_{i}J_{i}),
\end{equation}
where $f(x)$ is some function and $a_{i}$ is a constant. This means
that a natural choice of interpolation for a general orbit is
\begin{equation}
\mathcal{H}(L,J_{r}) = f(aL+bJ_{r}),
\label{eq:approx}
\end{equation}
where $a$ and $b$ are found using the solutions to Eqs. (\ref{eq:Jr})
and (\ref{eq:L}). Therefore, as long as $a$ and $b$ are available, one
has a very simple analytical approximation to the true Hamiltonian of
any scale-free spherical system.

If our Hamiltonian is not scale-free, we can often break the problem
up into scale-free regimes, and devise an interpolation formula that
smoothly matches the limits.  A specific example of this is provided
by \citet{Ev14}, who provided a Hamiltonian $\mathcal{H}(L,J_r)$ for a halo
model with a $1/r$ density cusp at the center but a flat rotation
curve at large radii.

\section{Power-Law Potentials}

\subsection{Results}
Let us now consider power-law potentials of the form
\begin{equation}
\Phi(r) = Ar^{\alpha},\qquad A = {v_0^2\over \alpha r_0^\alpha}.
\label{eq:pot}
\end{equation}
The choice of constant $A$ is motivated by the rotation law, which
takes the form
\begin{equation}
v_{\rm c}^2(r) = v_0^2\left({r\over r_0}\right)^{\alpha}.
\end{equation}
So, the rotation curve has the value $v_0$ at reference radius $r_0$.
Models with $\alpha >0$ (or $\alpha<0$) have a rising (or falling)
rotation curve. The model with $\alpha =0$ has a flat rotation curve,
and is recognised as the singular isothermal sphere. To recover its
familiar logarithmic potential, we must add a constant $-v_0^2/\alpha$
to eq.~(\ref{eq:pot}) and use L'H\^opital's rule to take carefully the
limit $\alpha \rightarrow 0$. The physical range of $\alpha$ is $-1\le
\alpha \le 2$, with the extremes representing a Keplerian potential
(point mass) and a harmonic potential (homogenous matter distribution)
respectively.

Applying our method to case when $A,\alpha > 0$ (the rising rotation
curve case), we find
\begin{eqnarray}
E(0,J_{r})&=&\bigg[\dfrac{\sqrt{2\pi}\alpha A^{1/\alpha}\Gamma(3/2+ 1/\alpha)}{\Gamma(1/\alpha)}\bigg]^{\beta}J_{r}^{\beta}, \nonumber \\
E(L,0)&=&\dfrac{(A\alpha)^{\beta/\alpha}}{\beta} L^\beta,
\end{eqnarray}
where we have set $\beta=2\alpha/(\alpha +2)$.  This implies that
$f(x)\propto x^{\beta}$. This gives us the following very simple
approximation for the Hamiltonian
\begin{equation}
\mathcal{H}(L,J_{r}) = C(L + DJ_{r})^{\beta},
\label{eq:crux}
\end{equation}
where the constants $C$ and $D$ are
\begin{eqnarray} \label{eq:powpos}
C &=& \dfrac{v_0^{2\beta/\alpha}}{r_0^\beta\beta}, \\
D &=&
\dfrac{\sqrt{2\pi}\Gamma(3/2+ 1/\alpha)\alpha^{-1/\alpha}\beta^{1/\beta}}{\Gamma(1+1/\alpha)}. \nonumber
\end{eqnarray}
This prescription may be tested on the special case of the harmonic
oscillator ($\alpha =2$). Setting $A=\frac{1}{2}\Omega^{2}$, we
recover the well-known exact analytical result (see, for example,
\citealt{Go80})
\begin{equation}
\mathcal{H}(L,J_{r}) = 2\Omega J_{r} + \Omega L.
\end{equation}
So, $\mathcal{H}$ perfectly coincides with the true Hamiltonian $H$
for the harmonic oscillator.

When $A,\alpha<0$ (the falling rotation curve case), the result is the
same as eq. (\ref{eq:crux}), except
\begin{equation}
D = \dfrac{\sqrt{2\pi}\Gamma(1-1/\alpha)(-\alpha)^{1-1/\alpha}(-\beta)^{1/\beta}}{\Gamma(-1/\alpha-1/2)}.
\end{equation}
The obvious test in this case is the Keplerian potential, where
$A=-GM$ and $\alpha=-1$, which gives:
\begin{equation}
\mathcal{H}(L,J_{r}) = -\dfrac{1}{2}\bigg(\dfrac{GM}{L+J_{r}}\bigg)^{2}.
\end{equation}
This coincides with the exact result for the Kepler potential, derived
by the residue calculus in \citet{Bo27} and repeated in \citet{Go80}.

The case of the flat rotation curve (or the singular isothermal
sphere) can be derived by taking the limit $\alpha \rightarrow 0$ and
using L'H\^opital's rule.  The potential becomes logarithmic
$\Phi=v_{0}^{2}\log r/r_0$, and so we find that $f(x) \propto
\log(x)$. Our approximation then gives
\begin{equation}
\mathcal{H}(L,J_{r}) = v_{0}^{2}\log\bigg(\dfrac{\sqrt{e}L+\sqrt{2\pi}J_{r}}{v_{0}r_0}\bigg),
\end{equation}
as stated in \citet{Ev14}.

So, we have found a surprisingly simple and compact
formula~(\ref{eq:crux}) for the Hamiltonian in terms of the actions
valid for any power-law potential with $-1\le \alpha \le2$. It is
exact for the two classical cases -- the Kepler potential and the
harmonic oscillator -- which have been known for over a
century. Notice too that the frequencies $\Omega_i$ are rationally
related only in the case of the Kepler potential and the harmonic
oscillator. Thus, we have recovered Bertrand's theorem~\cite[see
  e.g.,][]{Go80}, which states that the only central force laws with
everywhere closed bound orbits are the linear and inverse square force
laws. For all other cases, our formula~(\ref{eq:crux}) gives
frequencies that are incommensurable and so the motion is
conditionally periodic.

\subsection{Errors and First Order Corrections}

Given that the approximation is only exact for the cases $\alpha=-1$
and $\alpha=2$, we tested the effectiveness of the approximation for
several values of $\alpha$: $-0.5$, $0$ (the isothermal sphere),
$0.5$, $1$, and $1.5$. In each case we evaluate a set of 10000 orbits
on a grid of pericenter and apocenter radii: $r_{p}=0\rightarrow50$,
$r_{a}=0\rightarrow100$. For each orbit we evaluate the energy and
angular momentum, which are available as algebraic expressions in
terms of pericenter and apocenter radii \citep{Ly10}, and the radial
action, which requires the evaluation of a numerical integral. This
results in an irregularly spaced grid in action space.

Before discussing numerical values, it is necessary to explain how we
define errors across the sequence of models. We consider correction
terms of the form
\begin{equation} \label{eq:eps}
\mathcal{H} = f(aL+bJ_{r}+\epsilon\sqrt{LJ_{r}}).
\end{equation}
where $f(x) \propto x^\beta$ (if $\beta\neq0$) or $f(x) \propto \log
x$ (if $\beta =0$). Due to lack of dimensional constants in the
problem, the correction must possess the dimensions of actions. Its
form is suggested by examing the error distribution along rays in
action space.  A measure of the fractional error between the corrected
values and the original approximation is then
\begin{equation} \label{eq:errs}
\dfrac{\Delta\mathcal{H}^{1/\beta}}{\mathcal{H}^{1/\beta}}
=\dfrac{\epsilon\sqrt{LJ_{r}}}{aL+bJ_{r}+\epsilon \sqrt{LJ_r}}
\approx \dfrac{\epsilon\sqrt{LJ_{r}}}{aL+bJ_{r}},
\end{equation}
In the isothermal case ($\beta=0$), this becomes
\begin{equation}\label{eq:errsiso}
\dfrac{\Delta\mathcal{\exp H_{\mathrm{iso}}}}
{\exp \mathcal{H}_{\mathrm{iso}}}
\approx \dfrac{\epsilon\sqrt{LJ_{r}}}{aL+bJ_{r}}. 
\end{equation}
and so the measure of error is a continuous function of power-law
index $\alpha$.

\begin{figure}
	\centering
	\includegraphics[width=3in]{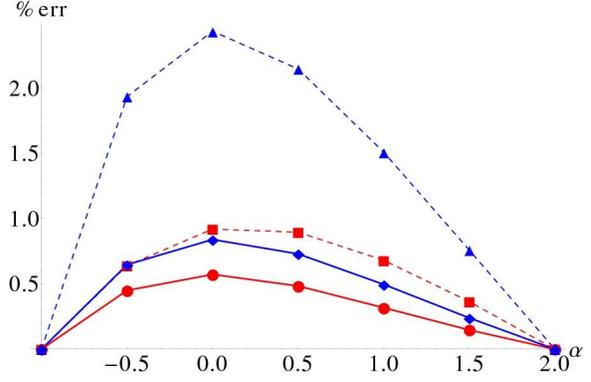}
\caption{Absolute percentage errors against power-law index
  ($\alpha=0$ corresponds to the isothermal sphere). The dashed lines
  correspond to the original approximation, and the solid lines
  include first-order corrections. Blue lines are maximum absolute
  percentage errors across the action grid, red lines are the mean
  absolute percentage errors. One can see that the first-order
  corrections reduce the maximum error significantly, and also have a
  noticeable effect on the mean error.}
\label{fig:maxmeanfo}
\end{figure}

\begin{figure}
\hspace{-0.5cm}\includegraphics[width=4in]{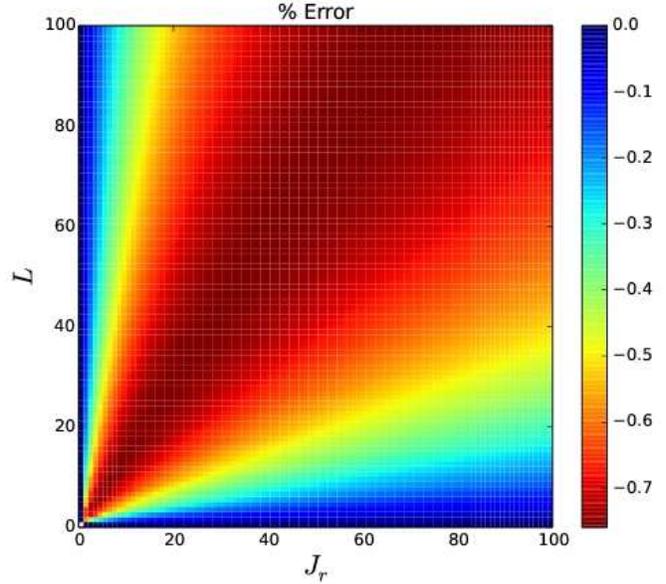}
\caption{Distribution of percentage error with actions for the case
  $\alpha=1.5$. One can see a definite systematic trend that could be
  removed with an astute choice of first order correction.}
\label{fig:errdist}
\end{figure}

We find that the errors in all cases are very small. The mean absolute
percentage error on the action grid is $<1$ percent in every case, and
the maximum absolute percentage error is always $<2.5$ percent. These
results are given in Fig. \ref{fig:maxmeanfo}. The distribution of
errors is very similar in all cases: it is scale-free and maximised
along a particular ray in action space. A typical example is given in
Fig. \ref{fig:errdist}.

Given the simplicity of the error distribution, we can calculate a
value for $\epsilon$ in Eq. (\ref{eq:eps}) very easily. Assuming the
errors are maximised along the line $L=\chi^2 J_{r}$, we can
substitute this into Eqs. (\ref{eq:errs}) or (\ref{eq:errsiso}) to
give
\begin{equation}
\epsilon = \dfrac{\delta_{\mathrm{max}}(\chi^2a+b)}{\chi}
\end{equation}
where $\delta_{\mathrm{max}}$ is the fractional error with the largest
absolute value on the grid. We carried out this analysis for each of
our test cases and found that it had a significant effect on the
maximum error on the grid, as well as reducing the average error in
every case. These results are also given in
Fig. \ref{fig:maxmeanfo}. If these first-order corrections are
included, the error never exceeds $1$ percent for any of our test
cases. 

On account of its astrophysical importance, we give the
expression for the first-order corrected formula for the
singular isothermal sphere:
\begin{eqnarray}
\mathcal{H} &=& v_{0}^2\log\bigg(\dfrac{\sqrt{e}L+\sqrt{2\pi}J_{r}+\epsilon\sqrt{LJ_{r}}}{v_{0}r_0}\bigg), \\
\epsilon &=& -0.1009. \nonumber 
\end{eqnarray}
For the case $\alpha=1$, which corresponds to the cosmologically
motivated $1/r$ or Navarro-Frenk-White (1996) density cusp, the
corrected expression is
\begin{eqnarray}
\mathcal{H} &=& {3v_0^{4/3}\over 2r_0^{2/3}} \bigg(L+ {\pi \over \sqrt{3}} J_{r}+\epsilon\sqrt{LJ_{r}}\bigg)^{2/3}, \\
\epsilon &=& -0.0412. \nonumber
\end{eqnarray}

\section{Applications}

Here, we present two simple applications of our results. The first is
to stellar streams. We demonstrate that within this approximation
streams are well delineated by orbits. The second is to cusps in the
centre of galaxies. We show that the central slope of a dark matter
halo is reduced when the stellar populations begin to dominate the
potential.

\subsection{Stellar Streams}

\citet{Sa13} showed that a stream can only be successfully modelled by
an orbit if the misalignment angle is very small. This angle is given
by
\begin{equation}
\varphi = \arccos(\hat{\boldsymbol{\Omega}}.\hat{\boldsymbol{e}}_{1}),
\end{equation}
where $\hat{\boldsymbol{\Omega}}$ is the normalised frequency vector
of the progenitor of the stellar stream and $\hat{\boldsymbol{e}}_{1}$
is the leading eigenvector of the Hessian matrix
\begin{equation}
D_{ij}(\hat{J})=\frac{\partial^{2}H}{\partial J_{i}\partial J_{j}}.
\end{equation}
The misalignment angle quantifies the offset in angle space between
stripped material and the progenitor, and encodes the direction in
which stars spread in action-angle space. For a long, thin stream to
form, the Hessian must have one dominant eigenvalue. We now evaluate
$D_{ij}$ within our approximate framework. Any approximate Hamiltonian
produced by our method is of the form
\begin{equation}
\mathcal{H}(L,J_{r})=f(aL+bJ_{r}),
\end{equation}
giving rise to a Hessian matrix
\begin{equation}
D_{ij}(\hat{J}) = G(L,J_{r})\left(\begin{array}{cc}
a^{2} & ab\\
ab & b^{2}\end{array}\right).
\end{equation}
Such a matrix has only one non-zero eigenvalue, $\lambda =
G(a^2+b^2)$, with corresponding eigenvector
\begin{equation}
\hat{\boldsymbol{e}}_{1}=\dfrac{1}{\sqrt{a^2/b^2+1}}\left(\begin{array}{c}
a/b\\
1\end{array}\right).\
\end{equation}
Now, if we evaluate a general frequency vector using $\mathcal{H}$, we find
\begin{equation}
\hat{\boldsymbol{\Omega}}\propto\hat{\boldsymbol{e}}_{1}.
\end{equation}
This implies that the misalignment angle vanishes, and so the stream
will always develop along an orbit under this approximation. This
suggests that, at least in the scale-free spherical case, the
misalignment angle must be very small and therefore the stream should
be very well approximated by an orbit.  This may provide a measure of
justification for the assumption used in \citet{Ko10}, in which the
GD1 stream~\citep{Gr06} was assumed to lie along an orbit in a
scale-free, mildly axisymmetric potential.


\subsection{Adiabatic Transformation of Density Cusps}

\citet*{Na96} famously discovered that, in dissipationless
simulations, dark haloes form a $1/r$ density cusp in the centre of
galaxies. However, there is a discrepancy between the predicted
density of dark matter at the center of spiral galaxies and the
inferred dark matter density consistent with observations. For
example, in the case of the Milky Way, \citet{Bi01} show that cuspy
dark matter density haloes would violate constraints on the rotation
curve, given the known contributions from the stellar and ISM disks
and Galactic bar. Clearly, the $1/r$ cusp, if it originally formed in
the Milky Way galaxy, has been weakened with the passage of time.

We assume that the dark halo of the galaxy formed first, and that the
dark matter density in the center of a galaxy originally followed a
$1/r$ cusp. With the build-up of the disk and bulge stellar
populations, we assume that the the density profile of the dominant
stellar populations becomes roughly constant in the very central
parts.  In this instance, the potential is initially linear ($\alpha
_{0}=1$), corresponding to the Navarro-Frenk-White cusp, and becomes
harmonic, ($\alpha _{1}=2$), corresponding to a constant density
core. Assuming that this process occurs sufficiently slowly that the
adiabatic invariance of the actions holds good, then we can find the
final profile of the dark matter.

For the collisionless dark matter, we use the self-consistent
power-law DF derived by \citet{Ev94} (Eq. 5.4), and assume an
initially isotropic population so that
\begin{equation}
f(E) = \mathcal{N}E^{- \dfrac{\alpha _{0} + 4}{2\alpha _{0}}},
\end{equation}
where $\mathcal{N}$ is a normalisation factor. We then insert our
approximation for $H(L,J_{r})$ from Eq. (\ref{eq:powpos}) for the case
$\alpha = \alpha _{0} = 1$ to find
\begin{equation}
f(L,J_{r}) \simeq \mathcal{M}\bigg(L+\dfrac{\pi J_{r}}{\sqrt{3}}\bigg)^{-5/3},
\end{equation} 
where $\mathcal{M}$ is a new normalisation factor. Since the changes
in the potential are presumed to occur adiabatically, the distribution
function is always of this form, even at the endpoint. We now
eliminate $J_{r}$ by again using Eq. (\ref{eq:powpos}) but for the
final potential, $\Phi_{1}=\frac{1}{2}\Omega^{2}r^2$ (which is exact
in this case), giving:
\begin{equation}
f(L,J_{r})\propto \bigg(L +
\dfrac{\pi}{2\sqrt{3}}\bigg[\dfrac{H}{\Omega} -
  L\big)\bigg]\bigg)^{-5/3}
\end{equation}
and substitute for $H$ using Eq. (\ref{eq:hxv}), which finally leaves
\begin{equation}
f(L,J_{r})\propto \bigg(\big(1-\dfrac{\pi}{2\sqrt{3}}\big)L +
\dfrac{\pi}{2\Omega\sqrt{3}}\big(\frac{1}{2}\boldsymbol{v}^{2} +
\frac{1}{2}\Omega^{2}r^{2}\big)\bigg)^{-5/3}.
\end{equation}  
We can simplify this rather messy looking expression. The first term
is proportional to $L$, which will be small for particles that
penetrate the centre of the galaxy. This term also has a considerably
smaller numerical prefactor than the isotropic term. On this basis, we
consider only the second term and integrate over velocity space to
find the spatial dependence of the final density profile
\begin{equation}
\rho_{\mathrm{dm}}(r) \propto \int_{0}^{\infty} \dfrac{v^2\mathrm{d}v}{\bigg(v^{2} + \Omega^{2}r^{2}\bigg)^{5/3}}.
\end{equation}
This integral can be evaluated using the substitution $v^{2} =
\Omega^{2}r^{2}\tan^{2}\theta$, and gives the final result
\begin{equation}
\rho_{\mathrm{dm}}(r) \propto r^{-1/3}.
\label{eq:nfwcase}
\end{equation}
So, we have found that the density profile of dark matter particles
that initially followed a $r^{-1}$ density cusp has developed a
shallower density profile of $r^{-1/3}$ as the density becomes
homogeneous in the very central parts.

This calculation can be carried out completely generally. Suppose the
initial dark matter cusp has form $r^{-\gamma_0}$. Let the
gravitational potential evolve so that the final end-state has $\phi
\propto r^{\alpha_{1}}$. Then the final dark matter density cusp
behaves like $r^{-\gamma_1}$, where
\begin{equation}
\gamma_1(\gamma_0,\alpha_1) = {\gamma_0 - \alpha_1(\gamma_0-3) -6\over \gamma_0 -4}.
\end{equation}
Our earlier result (\ref{eq:nfwcase}) corresponds to the special case
of an NFW cusp responding to a potential that slowly changes to
harmonic, that is $\gamma_1(1,2) = 1/3$

\begin{figure}
\hspace{-0.5cm}\includegraphics[width=3.5in]{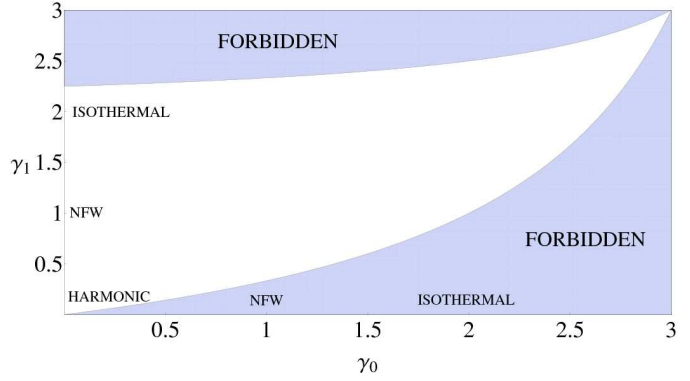}
\caption{The domain of allowed cusp slope transformations. The final
  cusp slope $\gamma_1$ is plotted against the initial cusp slope
  $\gamma_0$. Shaded regions cannot be reached by adiabatic
  changes. The upper line corresponds to $\gamma_1(\gamma_0,-1)$ (the
  Keplerian limit) and the lower line to $\gamma_1(\gamma_0,2)$ (the
  harmonic limit).}
\label{fig:domain}
\end{figure}

This equation enables us to map out the domain of accessible adiabatic
changes of cusp slope, as shown in Fig.~\ref{fig:domain}. The upper
line corresponds to $\gamma_1(\gamma_0,-1)$, in which the underlying
potential is slowly changed to Keplerian and the matter distribution
becomes strongly concentrated. The lower line corresponds to
$\gamma_1(\gamma_0,2)$, as the potential is changed to harmonic and
the matter distribution becomes homoegeneous. The starting cusp
$\gamma_0$ may be adiabatically transformed into $\gamma_1$ provided
\begin{equation}
{\gamma_0 \over 4 - \gamma_0} \le \gamma_1 \le {9- 2\gamma_0 \over
  4-\gamma_0}
\end{equation}
This region is shown unshaded in Fig.~\ref{fig:domain}. It is the
domain of accessible cusp transformations. Notice that an initial NFW
cusp can only ever be adiabatically transformed into a cusp with slope
$1/3 \le \gamma_1 \le 7/3$. It can never be completely erased.  Of
course, this result is restricted to adiabatic changes. There are a
variety of non-adiabatic processes that are widely believed to remove
cusps, including repeated mass loss~\citep{Re05}, AGN driven
outflows~\citep{Ma13} and infall of satellites~\citep{El01}.

Finally, we note that \citet{Go99}, in their study of the response of
a dark matter cusp to a growing black hole, derived the limit
$\gamma_1(\gamma_0,-1)$, though only for models with $0< \gamma_0
<2$. The formula is generally valid throughout the entire physical
range $0 \le \gamma_0 \le 3$, though the methods employed by
\citet{Go99} were unable to include isothermal cusp slopes and
steeper.

\section{Conclusions}

Action-angle coordinates are a powerful tool in modern classical
mechanics ~\citep[see e.g.,][]{Ab78, Ar89} and galactic
dynamics~\citep{Bi08}. Their widespread use is hampered by the fact
that the actions can be difficult to find explicitly. Although
numerical tools are becoming available to make this job
easier~\citep{Bi12,Bo14,Sa14}, there are disappointingly few
potentials for which analytic results are possible.

The main result of the paper is the discovery of a simple formula for
the Hamiltonian of scale-free potentials in terms of the actions. The
formula recovers the well-known and exact results for the harmonic
oscillator and Keplerian potentials, which have been known for over a
century~\cite[see e.g.,][]{Go80}. For other power-laws, the formula is
exact for radial and circular orbits, but approximate for general
orbits. However, numerical tests show that the the errors are always
very small, with mean errors across a grid of actions always $<1$
percent, and maximum errors never larger than $2.5$ percent. Our
formula therefore provides a useful and accurate approximate
Hamiltonian for many of the commonly used galactic potentials,
including the isothermal sphere and power-law cusps. It may also be
improved with simple first-order corrections if need be, which reduce
the mean error to always $<0.6$ percent, and the maximum error $<1$
percent.

We provided two applications. First, the mismatch of streams from
their progenitor is described by a misalignment angle. This in turn is
controlled by the Hessian of the Hamiltonian with respect to the
actions~\citep{Sa13}. Most streams in the Milky Way halo occur at
Galactocentric radii greater than 20 kpc, so that the potential is
probably scale-free to a good approximation. If, in addition, the
potential is nearly spherical, then the misalignment angle may be
close to zero. This suggests that there may be regimes in which the
approximation of thin streams by orbits is a reasonable one.

Second, we considered the evolution of a dark matter $1/r$ density
cusp in a spiral galaxy. Although $1/r$ density cusps are known to be
the endpoint of dissipationless simulations, there is ample evidence
that in galaxies like the Milky Way, any such cusp must have been
modified at the center~\citep{Bi01}. The response of a $1/r$ cusp to
the slow build-up of a stellar bulge and disk can be computed using
our formula, under the assumption of adiabatic invariance. As the
overall potential becomes cored and harmonic, the dark matter density
cusp becomes shallower and is nearly erased. It began as a $1/r$ and
finished as $1/r^{1/3}$ density cusp. Considering the more general
problem of density cusps of arbitrary slope, we show that adiabatic
transformations can only weaken, and never remove, such cusps. A
density singularity -- once present -- requires some non-adiabatic
process to erase it.

In this paper, we only considered scale-free potentials. Nonetheless,
our methods can also be applied to spherical potentials that possess a
scale-length. This requires more work because we have to devise a
suitable interpolation scheme to sew the different regimes together,
but it is still very effective. \citet{Ev14} provide a specific
example, but it would be interesting to study this problem in greater
generality.

Also looking to the future, it is worthwhile to consider whether
similar reasoning could be applied to scale-free, axisymmetric
potentials.  The regularity of Poincar\'e surfaces of sections for
axisymmetric scale-free potentials ~\citep{Ri82,Ev94} strongly hints
that there are simple results to be found.  This is a more complex
problem, given that there are now three actions to consider rather
than the two in the spherical case ($J_r$ and $J_\theta+J_\phi$).
Nonetheless, we can consider three special orbits: purely vertical
orbits, purely radial orbits and circular orbits in the plane, for
which the actions are well-defined. For example, the flattened
logarithmic potential is
\begin{equation}
\Phi(R,z) = \dfrac{1}{2}v_{0}^2\log\bigg((R/r_{0})^2 + (z/qr_{0})^2\bigg).
\end{equation}
Using the same interpolation techniques, we reach the following
expression for an approximate Hamiltonian:
\begin{equation}
\mathcal{H} = v_{0}^2\log\bigg(\dfrac{\sqrt{2\pi}(J_{R}+J_{z}/q)+\sqrt{e}|L_{z}|}{v_{0}r_{0}}\bigg).
\end{equation}
Given the fact that the actions cannot be written down as a
quadratures for general orbits in this potential, this expression
would have to be tested using an approximate method, such as described
in \citet{Bi12}. We look to do exactly this in future investigations.

\section*{Acknowledgments}

We thank Donald Lynden-Bell and Simon Gibbons for some extremely
useful conversations and insights. AW and AB acknowledge the support
of STFC.

\bibliography{approxh}

\begin{thebibliography}{}

\bibitem[\protect\citeauthoryear{{Abraham} \& {Marsden}}{{Abraham} \&
  {Marsden}}{1978}]{Ab78}
{Abraham} R.,  {Marsden} J.,  1978, {Foundations of Mechanics}.
Benjamin Cumings

\bibitem[\protect\citeauthoryear{{Agnello}, {Evans}, {Romanowsky} \&
  {Brodie}}{{Agnello} et~al.}{2014}]{Ag14}
{Agnello} A.,  {Evans} N.~W.,  {Romanowsky} A.~J.,    {Brodie} J.-P.,  2014,
  ArXiv e-prints

\bibitem[\protect\citeauthoryear{{Arnold}}{{Arnold}}{1989}]{Ar89}
{Arnold} V.,  1989, {Mathematical Methods of Classical Mechanics}.
Springer Verlag

\bibitem[\protect\citeauthoryear{{Binney}}{{Binney}}{2012}]{Bi12}
{Binney} J.,  2012, \mnras, 426, 1324

\bibitem[\protect\citeauthoryear{{Binney} \& {Tremaine}}{{Binney} \&
  {Tremaine}}{2008}]{Bi08}
{Binney} J.,  {Tremaine} S.,  2008, {Galactic Dynamics: Second Edition}.
Princeton University Press

\bibitem[\protect\citeauthoryear{{Binney} \& {Evans}}{{Binney} \&
  {Evans}}{2001}]{Bi01}
{Binney} J.~J.,  {Evans} N.~W.,  2001, \mnras, 327, L27

\bibitem[\protect\citeauthoryear{{Born}}{{Born}}{1927}]{Bo27}
{Born} M.,  1927, {The Mechanics of the Atom}.
G. Bell and Sons

\bibitem[\protect\citeauthoryear{{Bovy}}{{Bovy}}{2014}]{Bo14}
{Bovy} J.,  2014, ArXiv e-prints

\bibitem[\protect\citeauthoryear{{El-Zant}, {Shlosman} \& {Hoffman}}{{El-Zant}
  et~al.}{2001}]{El01}
{El-Zant} A.,  {Shlosman} I.,    {Hoffman} Y.,  2001, \apj, 560, 636

\bibitem[\protect\citeauthoryear{{Evans}}{{Evans}}{1994}]{Ev94}
{Evans} N.~W.,  1994, \mnras, 267, 333

\bibitem[\protect\citeauthoryear{{Evans}, {de Zeeuw} \& {Lynden-Bell}}{{Evans}
  et~al.}{1990}]{Ev90}
{Evans} N.~W.,  {de Zeeuw} P.~T.,    {Lynden-Bell} D.,  1990, \mnras, 244, 111

\bibitem[\protect\citeauthoryear{{Evans} \& {Williams}}{{Evans} \&
  {Williams}}{2014}]{Ev14}
{Evans} N.~W.,  {Williams} A.~A.,  2014, ArXiv e-prints

\bibitem[\protect\citeauthoryear{{Goldstein}}{{Goldstein}}{1980}]{Go80}
{Goldstein} H.,  1980, {Classical Mechanics: Second Edition}.
Addison-Wesley

\bibitem[\protect\citeauthoryear{{Gondolo} \& {Silk}}{{Gondolo} \&
  {Silk}}{1999}]{Go99}
{Gondolo} P.,  {Silk} J.,  1999, Physical Review Letters, 83, 1719

\bibitem[\protect\citeauthoryear{{Goodman} \& {Binney}}{{Goodman} \&
  {Binney}}{1984}]{Go84}
{Goodman} J.,  {Binney} J.,  1984, \mnras, 207, 511

\bibitem[\protect\citeauthoryear{{Grillmair} \& {Dionatos}}{{Grillmair} \&
  {Dionatos}}{2006}]{Gr06}
{Grillmair} C.~J.,  {Dionatos} O.,  2006, \apjl, 643, L17

\bibitem[\protect\citeauthoryear{{Katz}, {McGaugh}, {Sellwood} \& {de
  Blok}}{{Katz} et~al.}{2014}]{Ka14}
{Katz} H.,  {McGaugh} S.~S.,  {Sellwood} J.~A.,    {de Blok} W.~J.~G.,  2014,
  \mnras, 439, 1897

\bibitem[\protect\citeauthoryear{{Koposov}, {Rix} \& {Hogg}}{{Koposov}
  et~al.}{2010}]{Ko10}
{Koposov} S.~E.,  {Rix} H.-W.,    {Hogg} D.~W.,  2010, \apj, 712, 260

\bibitem[\protect\citeauthoryear{{Lynden-Bell}}{{Lynden-Bell}}{2010}]{Ly10}
{Lynden-Bell} D.,  2010, \mnras, 402, 1937

\bibitem[\protect\citeauthoryear{{Martizzi}, {Teyssier} \& {Moore}}{{Martizzi}
  et~al.}{2013}]{Ma13}
{Martizzi} D.,  {Teyssier} R.,    {Moore} B.,  2013, \mnras, 432, 1947

\bibitem[\protect\citeauthoryear{{Navarro}, {Frenk} \& {White}}{{Navarro}
  et~al.}{1996}]{Na96}
{Navarro} J.~F.,  {Frenk} C.~S.,    {White} S.~D.~M.,  1996, \apj, 462, 563

\bibitem[\protect\citeauthoryear{{Read} \& {Gilmore}}{{Read} \&
  {Gilmore}}{2005}]{Re05}
{Read} J.~I.,  {Gilmore} G.,  2005, \mnras, 356, 107

\bibitem[\protect\citeauthoryear{{Richstone}}{{Richstone}}{1982}]{Ri82}
{Richstone} D.~O.,  1982, \apj, 252, 496

\bibitem[\protect\citeauthoryear{{Sanders} \& {Binney}}{{Sanders} \&
  {Binney}}{2013}]{Sa13}
{Sanders} J.~L.,  {Binney} J.,  2013, \mnras, 433, 1813

\bibitem[\protect\citeauthoryear{{Sanders} \& {Binney}}{{Sanders} \&
  {Binney}}{2014}]{Sa14}
{Sanders} J.~L.,  {Binney} J.,  2014, ArXiv e-prints

\end{thebibliography}
\bibliographystyle{mn2e}

\end{document}